\documentclass[conference]{IEEEtran}
\IEEEoverridecommandlockouts

\usepackage{cite,comment}
\usepackage{mathtools}
\usepackage{hyperref}
\usepackage{graphicx}
\usepackage{enumitem}
\usepackage{tikz}
\usepackage{bm}
\usepackage{amsmath}
\usepackage{algorithm}
\usepackage{algorithmic}
\usepackage{amsmath,amssymb,amsthm}
\usepackage{array}
\usepackage{relsize}
\usepackage{subcaption}
\usepackage{mathtools, nccmath}
\usepackage[labelformat=simple]{subcaption}
\usepackage{derivative}
\usepackage{cuted}     
\usepackage{afterpage}
\usepackage{lipsum}
\usepackage[thinc]{esdiff}

\usepackage[top    = 0.80in,
bottom = 1.05in,
left   = 0.80in,
right  = 0.80in]{geometry}

\newcolumntype{P}[1]{>{\centering\arraybackslash}p{#1}}
\DeclareMathOperator{\C}{\mathbb{C}}

\DeclareMathOperator*{\argmax}{arg\,max}

\DeclarePairedDelimiter\abs{\lvert}{\rvert}
\DeclareCaptionSubType * [alph]{table}

\newcommand{\h}{\mathbf{H}}

\newcommand{\I}{\mathbf{I}}

\newcommand{\U}{\mathbf{U}}

\newcommand{\as}{\mathbf{a}}
\newcommand{\bs}{\mathbf{b}}

\newcommand{\js}{{\rm j}}

\newcommand{\vs}{\mathbf{v}}

\newcommand{\hs}{\mathbf{h}}

\newcommand{\xs}{\mathbf{x}}

\newcommand{\zs}{\mathbf{z}}

\newcommand{\hr}{\mathsf{H}}
\newcommand{\tr}{\mathsf{T}}

\newcommand{\bc}{\begin{center}}
\newcommand{\ec}{\end{center}}

\newcommand{\ds}{\displaystyle}
\newcommand{\uprightsubscript}[1]{_{\textnormal{#1}}}
\begingroup\lccode`~=`?\lowercase{\endgroup\let~}\uprightsubscript
\AtBeginDocument{\mathcode`?="8000 }

\newcommand{\tn}[1]{\textnormal{#1}}

\renewcommand{\baselinestretch}{1.0}
\theoremstyle{remark}

\definecolor{dg}{RGB}{255,0,0}

\begin{document}

\title{ Joint Communication and Sensing in RIS-Assisted MIMO System Under Mutual Coupling
}

\author{Dilki Wijekoon, Amine Mezghani, and Ekram Hossain,~\IEEEmembership{Fellow,~IEEE}

\thanks{The authors are with the Department of Electrical and Computer Engineering, University of Manitoba, Winnipeg, Manitoba, Canada (e-mails:  wijekood@myumanitoba.ca, {amine.mezghani, ekram.hossain\}@umanitoba.ca).} This work was supported in part by Technology Innovation Institute (TII), Abu Dhabi, UAE.}
}

\maketitle

\begin{abstract}
This paper considers a downlink Reconfigurable Intelligent Surface (RIS)-assisted Joint Communication and Sensing (JCAS) system within a physically-consistent setting, accounting for the effect of mutual coupling between RIS elements arising due to sub-element spacing. The system features a multiple-input multiple-output (MIMO) terrestrial base station (BS) and explores both monostatic and bistatic radar configurations to enable joint communication and sensing. In the monostatic configuration, both the transmitter and receiver are at the same location, while the bistatic configuration separates the transmitter and receiver spatially. System performance is evaluated using Fisher Information (FI) to quantify sensing accuracy and Mutual Information (MI) to measure communication efficiency. To achieve an optimal balance between communication and sensing, the RIS reflective coefficients and BS transmit beamforming are jointly optimized by maximizing a weighted sum of FI and MI. A novel solution approach is proposed for a single-user, single-object scenario, leveraging the mutual coupling model to enhance system realism. The impact of self-interference on sensing performance is also investigated through signal quantization. Numerical results reveal a fundamental trade-off between FI and MI and demonstrate that incorporating mutual coupling within a physically-consistent framework significantly improves both communication and sensing performance compared to conventional RIS-assisted JCAS models. Additionally, the analysis highlights how the choice of monostatic versus bistatic radar configuration affects system performance, offering valuable insights for the design of RIS-assisted JCAS systems.
\end{abstract}
\vspace{3pt}
\begin{IEEEkeywords}
Reconﬁgurable Intelligent Surface (RIS), Joint Communication and Sensing (JCAS), monostatic radar, bistatic radar, mutual coupling, mutual information, fisher information, physically-consistent modeling, self interference, quantization. 
\end{IEEEkeywords}

\section{Introduction} 
\subsection{Background}
The rapid growth in wireless communication technologies has greatly increased the demand for frequency spectrum, making efficient spectrum allocation a critical issue due to the limited availability of the spectrum \cite{verde2024integratingsensingcommunicationssimultaneously,article,10052711,inproceedings, RISsurvey2023, 8811733, 9721205, 8910627,9174801}. Joint Communication and Sensing (JCAS) has become an emerging technique due to its ability to optimize resource utilization, reduce system complexity, and support emerging applications such as autonomous vehicles, smart cities, and Internet of Things (IoT) networks. 
JCAS addresses spectrum-sharing challenges by enabling a shared platform to support both communication and sensing functionalities simultaneously.

One major application of JCAS is beamforming, leveraging the increased spatial diversity offered by MIMO systems \cite{CRB_article,inproceedings,10615441}. In \cite{CRB_article,inproceedings}, the authors present a design for joint beamforming in Joint Radar Communication (JRC) that takes into account the Cramér-Rao Lower Bound (CRLB) corresponding to the radar sensing and Signal-to-Interference-plus-Noise Ratio (SINR) for communication users as the performance metrics. The work in \cite{10615441} presents a joint beamforming method for JCAS considering a terrestrial MIMO BS that utilizes Fisher Information (FI) related to the sensing object and Mutual Information (MI) for communication users. This approach aims to achieve the FI/MI Pareto boundary, optimizing the trade-off between sensing accuracy and communication performance. Unlike CRLB, FI directly relates to the statistical efficiency of estimators and can dynamically adapt to changes in the system, making it particularly suitable for JCAS. The FI allows for joint optimization of sensing and communication by providing a tractable way to balance these two functions, especially when combined with MI for communication performance. 

Experiments that focus solely on base station (BS)-based JACS restrict the available Degrees of Freedom (DoF) for JCAS, limiting its flexibility and effectiveness in adapting to dynamic environments. The BSs have a fixed number of antennas and are often limited by hardware constraints and deployment environments, which restrict their beamforming flexibility. Consequently, JCAS systems relying solely on the BSs provide low-resolution performance for both sensing and communication.
However, these limitations can be overcome by integrating Reconfigurable Intelligent Surfaces (RIS) into the system. 
RIS consists of a planar array of passive, low-cost elements whose reflective properties, such as phase shifts and amplitudes, can be dynamically adjusted via external control signals \cite{ABoI_EURASIP,RISsurvey2023,RIS_ISAC_SPM}. By adjusting the reflected signals in real time, RIS enhances beamforming capabilities, offering additional degrees of freedom (DoF) that improve spatial resolution, extend coverage, and enable more precise joint communication and sensing functionalities. In JCAS systems, RIS enhances both communication and sensing links by creating virtual line-of-sight paths, and focusing reflected signals to improve sensing accuracy. Several works explore RIS-assisted JCAS frameworks showing that RIS enables significant improvements in both communication and sensing simultaneously \cite{10052711,10364760,10197455}.

Based on the geometrical configuration of the transmitter and receiver at the BS, JCAS can be categorized into two configurations. They are called monostatic and bistatic radar configurations \cite{7485181,10065107,845248}. Monostatic radar involves co-located transmitters and receivers, where the transmitted signal reflects off a target and returns to the same location for processing \cite{sheemar2024duplexjointcommunicationssensing, 10403515, 10438390, 6159700, luo20236dradarsensingtracking}. This configuration offers simplicity, high-resolution sensing for close-range scenarios, and reduced synchronization challenges. However, it suffers from limited coverage, increased interference, and an inability to effectively capture targets in NLoS environments.
In contrast, bistatic radar spatially separates the transmitter and receiver, allowing for flexible deployment and broader coverage \cite{4201772, 4720719, 6159691,10486914,10946614, 10788035}. This configuration excels at detecting targets, leveraging multipath propagation, and reducing interference. Bistatic systems can also utilize RIS to create customized propagation paths, enhancing both communication and sensing in NLoS scenarios. In a monostatic radar configuration, self-interference can be significant due to the close proximity of the transmit and receive antennas. This can be degraded system performance, hence self-interference mitigation plays a crucial role. Consequently, many studies have employed various self-interference suppression and elimination techniques in their experiments \cite{9731637,10325195,9275349,10890380,10121669,7222458,9048786,9443340,9376319}.

In the context of RIS, mutual coupling is unavoidable due to the sub-wavelength structure of the reflective elements.
Arranging the elements closer strengthens mutual coupling, and adding more elements to the RIS further amplifies the coupling effects \cite{10096563,9360851,10417011,wijekoon2024physicallyconsistentmodelingoptimizationnonlocal}. The majority of existing works characterize mutual coupling using impedance parameters ($Z$-parameters) under the minimum scattering assumption, typically considering dipole-based antenna elements. Some studies have also employed scattering parameters ($S$-parameters) to describe mutual coupling. All these studies confirm that leveraging mutual coupling in RIS-based optimizations leads to enhanced system performance compared to conventional RIS models that neglect mutual coupling. Although all these studies utilize RIS mutual coupling to enhance communication performances, none of these works have explored its impact on the performance of joint communication and sensing (JCAS).

\subsection{Contributions}

Motivated by the considerations described above, we propose a novel method for optimizing joint transmit beamforming and reflection coefficients in an RIS-assisted JCAS system. Our approach takes into account the effects of mutual coupling among  RIS elements in a physically-consistent setting, offering a more realistic representation of RIS-assisted JCAS systems. We adopt a more general mutual coupling modeling framework that adheres to fundamental laws of physics, moving beyond the conventional dipole antenna-based mutual coupling approaches. This modeling allows for a more accurate and flexible representation of RIS element interactions, capturing the complex electromagnetic behaviors in practical deployment scenarios. Our aim is to demonstrate that incorporating the effects of mutual coupling into JCAS optimization significantly enhances system performance. 

The contributions of this paper can be summarized as follows:

\begin{itemize}
    \item We propose a JCAS framework that leverages MI and FI for the joint design of transmit beamforming for the sensing target and the communication user, as well as the reflection coefficients for the RIS in a physically-consistent setting considering mutual coupling.
    
    \item A detailed analysis is provided for both monostatic and bistatic radar configurations. For each setup, the joint design of transmit beamforming and RIS reflection coefficients is formulated while incorporating mutual coupling, capturing the unique system dynamics, and offering a comprehensive understanding of its impact on sensing and communication performance.
    
    \item A general mutual coupling model based on scattering parameters is employed, which extends the traditional dipole element-based impedance models.
    
    \item In the monostatic radar configuration, the effect of self-interference on sensing performance is investigated through quantization of the sensing signal \cite{ordin2024foundations,Mezghani2016,5456454}, rather than elimination or mitigation of this effect as assumed in the state-of-the-art methods, allowing direct observation of how self-interference influences FI for sensing.
    
    \item We develop an optimization algorithm that maximizes the weighted sum of FI for sensing and MI for communication, aiming to strike a balance between the two objectives. The proposed solution employs projected gradient ascent and utilizes MATLAB CVX for efficient optimization.
    
    \item We present optimization solutions for both conventional and physically consistent RIS models to enable a comparative analysis. The conventional model assumes ideal RIS elements, neglecting mutual coupling effects, and provides a simplified representation of RIS behavior.
    
    \item Simulation results are presented to explore the trade-off between MI and FI. The numerical analysis demonstrates that incorporating mutual coupling in RIS significantly enhances system performance, improving both sensing and communication capabilities across monostatic and bistatic radar configurations.
    
\end{itemize}

The rest of the paper is organized as follows: Section II outlines the system model and assumptions. Then, Section III describes the effect of self-interference. Section IV presents the  problem formulation, the proposed solution and its computational complexity.  Section V presents the numerical results and  Section VI concludes the paper. \\
\textbf{Notations:} Matrices are denoted by capital boldface letters (e.g., $\mathbf{A}$) and vectors are denoted by small boldface letters (e.g., $\mathbf{a}$). The operators $(.)^\tr$, $(.)^*$, and $(.)^\hr$ denote the transpose, conjugate and conjugate transpose. The trace is denoted by $\text{Tr}(\mathbf{A})$.

\section{System Model and Assumptions}

\begin{figure}[h!]
\centering
\includegraphics[scale=0.3]{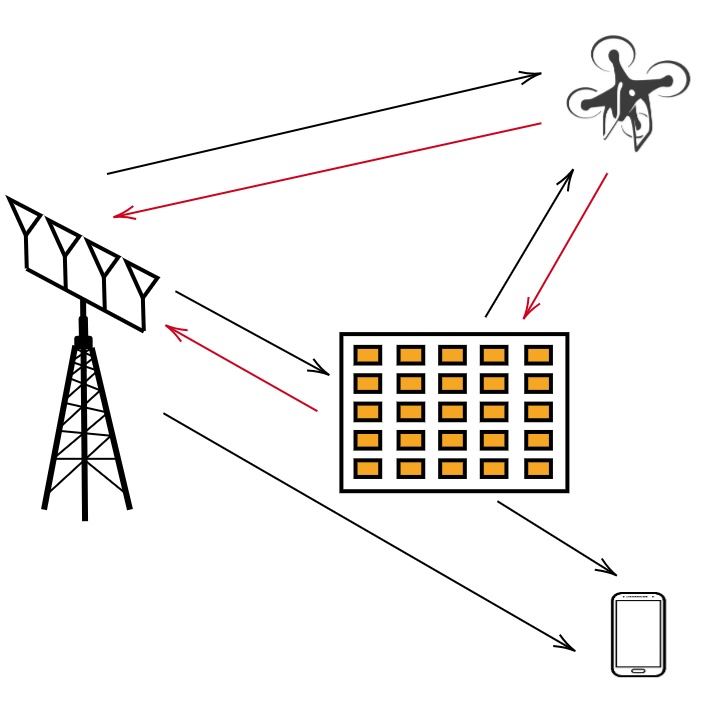}
\vspace{-2mm} 
\caption{System model of JCAS for monostatic radar.}
\label{fig:sys}
\end{figure}

\begin{figure}[h!]
\centering
\includegraphics[scale=0.3]{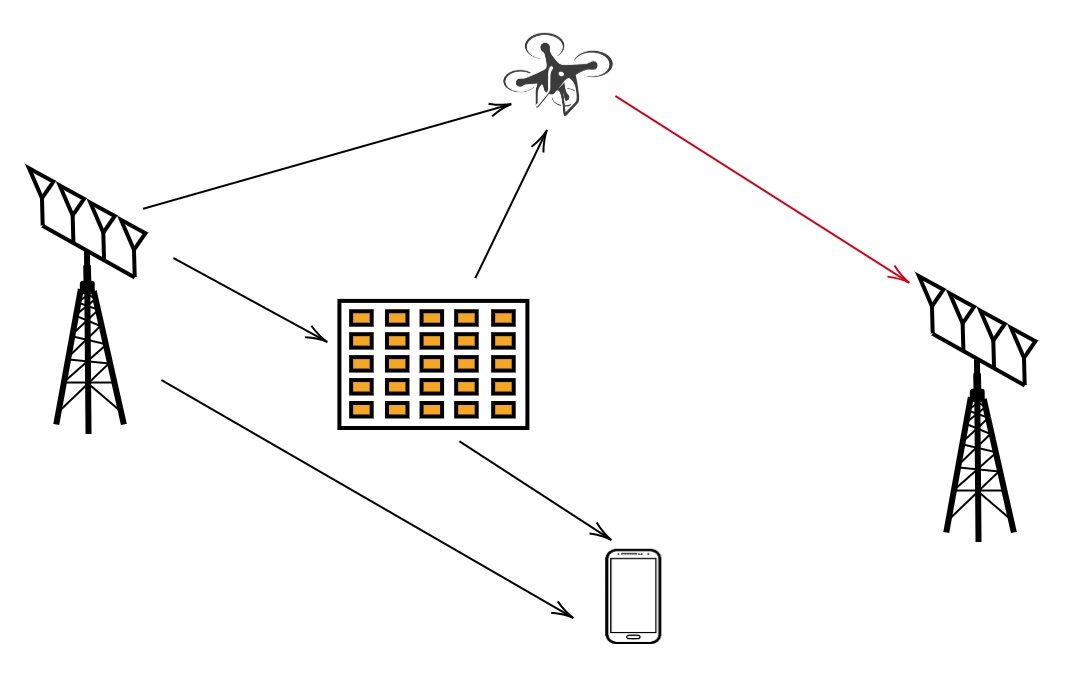}
\vspace{-2mm} 
\caption{System model of JCAS for bistatic radar.}
\label{fig:sys_bi}
\end{figure}

We consider an RIS-assisted downlink system with a MIMO terrestrial BS, analyzing monostatic and bistatic radar configurations separately. The system setups consist of a single RIS and a single user, as illustrated in Figures~\ref{fig:sys} and \ref{fig:sys_bi} for the monostatic and bistatic cases, respectively. In the monostatic configuration, the BS employs arrays of $N_t$ transmit and $N_r$ receive antennas that are co-located, whereas in the bistatic configuration, the $N_t$ transmit and $N_r$ receive antennas are positioned at different locations. In both cases, the RIS comprises of $M$ reflective elements.   
\vspace{-10pt}

\subsection{Communication Model}

Consider a JCAS signal from the BS, denoted by 
$\xs \in \C^{N_{t}}$ which is a Gaussian vector. The corresponding covariance matrix from the BS, $\mathbf{R_{\xs}} \in \C^{N_{t} \times N_{t}}$ is subject to the constraint $\text{Tr}(\mathbf{R_{\xs}}) \leq P$, where $P$ is the total transmit power \cite{10615441}. 
The total received signal at the communication user is: 
\begin{equation}
\label{eqn:eq1}
y_{c}=\underbrace{\hs?{r-u}^\hr\Theta\h?{b-r}\xs}?{BS-RIS-User}  + \underbrace{\hs?{b-u}^\hr\xs}?{BS-User} +\ z_{c}.
\end{equation}
Here, $\hs?{r-u}^\hr \in \C^{1 \times M}$ denotes the channel vector from the RIS to the user, while the matrix $\h?{b-r} \in \C^{M \times N_{t}}$ represents the channel matrix from BS to the RIS.  The vector $\hs?{b-u}^\hr \in \C^{1 \times N_t}$ represents the direct communication channel between the BS and the user. The scalar $\ z_{c} \in \C$ is the additive white Gaussian noise (AWGN) with zero mean and $\sigma_{c}^2$ variance. The matrix $\Theta \in \C^{M \times M}$ is the phase shift matrix. In the physically-consistent model, the phase shift matrix $\Theta$ is represented as a non-diagonal matrix, given by:
\begin{equation}
\label{eqn:eq2}
    \Theta={(\mathbf{\Upsilon}^{-1}-\mathbf{S})^{-1}},
\end{equation}
where $\mathbf{\Upsilon}=\tn{Diag}(\bm\upsilon) \in \C^{M \times M}$ is a diagonal matrix and $\bm\upsilon \in \C^M$ is the phase shift vector \cite{9474428,10417011,10096563}. The elements of phase shift vector $\bm\upsilon$ can be denoted by $\upsilon_m=e^{\js\theta_m}$, where $\abs*{\upsilon_m}=1$ for $m=1,\cdots, M$. The matrix  $\mathbf{S} \in \C^{M \times M}$ is the scattering matrix associated with mutual coupling effects. The following equations are used to construct the scattering matrix \cite{10417011,10096563}:
\begin{equation}
\label{eqn:sm1}
\mathbf{S}= \U\sqrt{\I_M-\Lambda}\U^\tr,
\end{equation}
\vspace{-15pt}
\begin{equation}
\label{eqn:sm2}
\mathbf{B}~=\I_M-\mathbf{S}\mathbf{S}^\hr~=\U\Lambda\U^\hr.
\end{equation}
The matrix $\mathbf{B}$, known as the embedded pattern coupling matrix, is constructed based on looseness, reciprocity, and cosine radiation patterns, as detailed in \cite{10417011,10096563}. Its structure depends on the RIS element number and spacing between elements. After determining $\mathbf{B}$
based on \cite{10417011,10096563}, the scattering matrix $\mathbf{S}$ can be obtained from equations (\ref{eqn:sm1}) and (\ref{eqn:sm2}). The goal is then to optimize the reflection coefficient matrix $\mathbf{\Upsilon}$, while accounting for the mutual coupling effects captured by $\mathbf{S}$.
In the absence of mutual coupling, i.e., $\mathbf{S}=\mathbf{0_{M}}$, which simplifies to the conventional model where the phase shift matrix becomes $\Theta=\mathbf{\Upsilon}$. 

We consider mutual information MI to analyze the performance of the communication model \cite{10615441}. Maximizing the MI at the user enables a substantial amount of the signal to be effectively delivered to the communication user. Thereby, the MI for vector $\xs$ is denoted by \cite{10615441}:
\begin{equation}
\label{eqn:eq3}
    \mathrm{MI}= \log_{2}\left(1+\frac{\hs^\hr\mathbf{R_{\xs}}\hs}{\sigma_{c}^2} \right),
\end{equation}
where $\hs^\hr \in \C^{1 \times N_t}$ is the total channel vector given by:
\begin{equation}
\label{eqn:eq4}
    \hs^\hr=\hs?{r-u}^\hr\Theta\h?{b-r}  + \hs?{b-u}^\hr.
\end{equation}
\subsection{Sensing Model for Monostatic Radar}

For the sensing object, the reflected signal is given by: 
\begin{equation}
\label{eqn:eq5}
\boldsymbol{y_{sm}}=\gamma_{2}\h?{r-b}\Theta^\tr\mathbf{A_{2}}\Theta\h?{b-r}\xs  + \gamma_{1}\mathbf{A_{1}}\xs +\boldsymbol{\zs_{sm}}.
\end{equation}
Here, $\h?{b-r} \in \C^{M \times N_{t}}$  is the channel matrix from the transmit antennas to RIS, and $\h?{r-b} \in \C^{M \times N_{r}}$ is the channel matrix from RIS to receive antennas. The matrix $\Theta$ is the phase shift matrix and  $\boldsymbol{\zs_{sm}} \in \C^{N_r}$ represents the additive white Gaussian noise (AWGN) with zero mean and variance 
$\sigma_{sm}^2$ per each. The scalars $\gamma_{1} \in \C$ and $\gamma_{2} \in \C$ represent the reflection coefficients, accounting for the round-trip path loss and radar cross-section for the BS-sensing object link and the RIS-sensing object link, respectively.
The matrix $\mathbf{A_{1}} \in \C^{N_{r} \times N_{t}}$  is given by:
\begin{equation}
\label{eqn:eq6}
    \mathbf{A_{1}}=\bs(\theta_{1})\as^\tr(\theta_{1}),
\end{equation}
where $\as(\theta_{1}) \in \C^{N_{t}}$, and $\bs(\theta_{1}) \in \C^{N_{r}}$ represent the transmit and receive array response vectors respectively corresponding to the BS, while matrix $\mathbf{A_{2}} \in \C^{M \times M}$ is given by:
\begin{equation}
\label{eqn:eq7}
\mathbf{A_{2}}=\bs(\theta_{2},\psi)\bs^\tr(\theta_{2},\psi),
\end{equation}
in which $\bs(\theta_{2},\psi) \in \C^{M}$ represents the array response vector corresponding to the RIS. The angles $\theta_{1}$ and $\theta_{2}$ denote the angle of the sensing object relative to the BS and RIS. Since this is a monostatic radar configuration, we assume that the Angle of Arrival (AOA) and the Angle of Departure (AOD) are the same. 

We consider the Fisher Information matrix to evaluate the performance of the sensing model \cite{10615441}. 
Based on the estimation vector $\vs=[\theta _{1} \ \gamma _{1} \ \gamma^*_{1} \ \theta _{2} \ \gamma _{2} \ \gamma^*_{2}]^\tr$ and considering the monostatic radar model given by equation (\ref{eqn:eq5}), the FI matrix $\mathbf{F}$ is given by:
 \begin{equation}
\label{eqn:eq8}
    \mathbf{F} = 
    \begin{bmatrix}
    F_{11}  & F_{12}  & F_{13}  & F_{14}  & \quad F_{15}  & F_{16} \\
    F_{21} &  F_{22} & 0 & F_{24} & F_{25} &  0 \\
    F_{31} & 0 & F_{33} & F_{34} & 0 & F_{36} \\
    F_{41} & F_{42} & F_{43} & F_{44} & F_{45} & F_{46} \\
    F_{51} & F_{52} & 0 & F_{54} & F_{55} & 0 \\
    F_{61} & 0 & F_{63} & F_{64} & 0 & F_{66} 
    \end{bmatrix}.
    \end{equation}
In this study, we only focus on the diagonal elements of the FI matrix, as the trace of the FI matrix is used for derivations and calculations in the subsequent sections. The diagonal elements of the FI matrix are expressed as:   
   \begin{equation}
   \label{eqn:eq9}
            F_{11} = \frac{2\gamma_{1}\gamma_{1}^*}{\sigma_{s}^2}\ \text{Tr}\left(\mathbf{\dot{A_1}}\mathbf{R_{\xs}}\mathbf{\dot{A^\hr_1}}\right)
\end{equation}
\begin{equation}
\label{eqn:eq10}
        F_{22} = \frac{1}{\sigma_{s}^2}\text{Tr}\left(\mathbf{A_1}\mathbf{R_{\xs}}\mathbf{A^\hr_1}\right)
        \end{equation}
        \begin{equation}
        \label{eqn:eq11}
            F_{33} = \frac{1}{\sigma_{s}^2}\text{Tr}\left(\mathbf{A_1}\mathbf{R_{\xs}}\mathbf{A^\hr_1}\right)
        \end{equation}
        \begin{align}
      \label{eqn:eq12}
       F_{44} = & \frac{2\gamma_{2}\gamma_{2}^*}{\sigma_{s}^2}\ \text{Tr}\left(\h?{r-b}\Theta^\tr\mathbf{\dot{A_2}}\Theta\h?{b-r}\mathbf{R_{\xs}}\h?{b-r}^\hr\right. \nonumber \\
       & \left.\Theta^\hr\mathbf{\dot{A^\hr_2}}(\h?{r-b}\Theta^\tr)^\hr\right)
     \end{align}
        \vspace{5pt}
        \begin{equation}
         \label{eqn:eq13}
            F_{55} = \frac{1}{\sigma_{s}^2}\text{Tr}\left(\h?{r-b}\Theta^\tr\mathbf{{A_2}}\Theta\h?{b-r}\mathbf{R_{\xs}}\h?{b-r}^\hr\Theta^\hr\mathbf{{A^\hr_2}}(\h?{r-b}\Theta^\tr)^\hr\right)
           \end{equation}
        \begin{equation}
        \label{eqn:eq14}
            F_{66} =  \frac{1}{\sigma_{s}^2}\text{Tr}\left(\h?{r-b}\Theta^\tr\mathbf{{A_2}}\Theta\h?{b-r}\mathbf{R_{\xs}}\h?{b-r}^\hr\Theta^\hr\mathbf{{A^\hr_2}}(\h?{r-b}\Theta^\tr)^\hr\right).
            \vspace{5pt}
        \end{equation}
Here, $\dot{A_1}=\frac{\partial\mathbf{A_{1}}}{\partial\theta_{1}}=\bs(\theta_1)\dot{\as}^\tr(\theta_1)+\dot{\bs}(\theta_1)\as^\tr(\theta_1)$ and $\dot{A_2}=\frac{\partial\mathbf{A_{2}}}{\partial\theta_{2}}=\bs(\theta_2,\psi)\dot{\bs}^\tr(\theta_2,\psi)+\dot{\bs}(\theta_2,\psi)\bs^\tr(\theta_2,\psi)$, where $\mathbf{A_{1}}$ and $\mathbf{A_{2}}$ are defined in equations (\ref{eqn:eq6}) and (\ref{eqn:eq7}), respectively. The derivative $\dot{\bs}(\theta_2,\psi)$ is the partial derivative w.r.t. $\theta_2$.
For the array response vectors 
 $\as(\theta_{1})$ and $\bs(\theta_{1})$  we consider a Uniform Linear Array (ULA) with the origin located at the center of the array \cite{10615441}. The array response vectors and their corresponding derivatives are given by \cite{10615441}:
\begin{align}
\label{eqn:e26}
&\nonumber
   \as(\theta_{1})=\bs(\theta_{1})=
       \\ & {\left[ {{e^{ - j\frac{{{N_t-1}}}{2} k \sin \theta_1 }},{e^{ - j\frac{{{N_t} - 3}}{2} k \sin \theta_1 }}, \ldots, {e^{j\frac{{{N_t-1}}}{2} k \sin \theta_1 }}} \right]^T}, 
\end{align} 
\vspace{-15pt}
\begin{align}
\label{eqn:e27}
&\nonumber
    \dot{\as}(\theta_{1})= \dot{\bs}(\theta_{1})= \\& {\left[ { - j{a_1}\frac{{{N_t-1}}}{2} k \cos \theta_1, \ldots, -j{a_{{N_t}}}\frac{{{N_t-1}}}{2} k \cos \theta_1 } \right]^T}.
\end{align}
Here, $k= \frac{2\pi d}{\lambda}$ with $d$ representing the distance between two antennas in the array, and $\lambda$ denoting the wavelength. Also, $a_i$ represents the $i$-th element of $\as(\theta_{1})$ \cite{10615441}. For the array response vector of the RIS $\bs(\theta_2,\psi)$, a square Uniform Planar Array (UPA) with a cosine radiation pattern is utilized.
The corresponding array response vector for the RIS, as described in \cite{9474428, 10417011}, is given by:
\begin{align}
\label{eqn:eq28}
 \begin{split}
\bs(\theta_2,\psi) = \sqrt{|\cos\theta_2|}&
\begin{bmatrix}
1\\
e^{2\pi \js\frac{d?r}{\lambda}\sin\theta_2\sin\psi}\\
\vdots\\
e^{2\pi \js\frac{d?r}{\lambda}(\sqrt{M}-1)\sin\theta_2\sin\psi}
\end{bmatrix} \otimes \\
&\begin{bmatrix}
1\\
e^{2\pi \js\frac{d?r}{\lambda}\sin\theta_2\cos\psi}\\
\vdots\\
e^{2\pi \js\frac{d?r}{\lambda}(\sqrt{M}-1)\sin\theta_2\cos\psi}
\end{bmatrix}.
    \end{split}
\end{align}
The scalar $d?r$ is the distance between two consecutive RIS elements, while the angle $\psi$ represents the angle of azimuth. The partial derivative of the vector w.r.t. $\theta_2$ is denoted as in the equation (\ref{eqn:eq29}).
\begin{table*}
\small
    \begin{align}
    \label{eqn:eq29}
    \dot{\bs}(\theta_2, \psi) &= \frac{-\sin\theta_2\cos\theta_2}{2|\cos\theta_2|^\frac{3}{2}}
    \begin{bmatrix}
    1 \\
    e^{2\pi j\frac{d_r}{\lambda} \sin \theta_2 \sin \psi} \\
    \vdots \\
    e^{2\pi j\frac{d_r}{\lambda} (\sqrt{M} - 1) \sin \theta_2 \sin \psi}
    \end{bmatrix} \otimes
    \begin{bmatrix}
    1 \\
    e^{2\pi j\frac{d_r}{\lambda} \sin \theta_2 \cos \psi} \\
    \vdots \\
    e^{2\pi j\frac{d_r}{\lambda} (\sqrt{M} - 1) \sin \theta_2 \cos \psi}
    \end{bmatrix} \notag \\[8pt]
    &\quad + \sqrt{|\cos \theta_2|}
    \begin{bmatrix}
    0 \\
    2\pi j\frac{d_r}{\lambda} \cos \theta_2 \sin \psi \, 
    e^{2\pi j\frac{d_r}{\lambda} \sin \theta_2 \sin \psi} \\
    \vdots \\
    2\pi j\frac{d_r}{\lambda} (\sqrt{M} - 1) \cos \theta_2 \sin \psi \, e^{2\pi j\frac{d_r}{\lambda} (\sqrt{M} - 1) \sin \theta_2 \sin \psi}
    \end{bmatrix} \otimes
    \begin{bmatrix}
    1 \\
    e^{2\pi j\frac{d_r}{\lambda} \sin \theta_2 \cos \psi} \\
    \vdots \\
    e^{2\pi j\frac{d_r}{\lambda} (\sqrt{M} - 1) \sin \theta_2 \cos \psi}
    \end{bmatrix} \notag \\[8pt]
    &\quad + \sqrt{|\cos \theta_2|}
    \begin{bmatrix}
    1 \\
    e^{2\pi j\frac{d_r}{\lambda} \sin \theta_2 \sin \psi} \\
    \vdots \\
    e^{2\pi j\frac{d_r}{\lambda} (\sqrt{M} - 1) \sin \theta_2 \sin \psi}
    \end{bmatrix} \otimes
    \begin{bmatrix}
    0 \\
    2\pi j\frac{d_r}{\lambda} \cos \theta_2 \cos \psi \,
    e^{2\pi j\frac{d_r}{\lambda} \sin \theta_2 \cos \psi} \\
    \vdots \\
    2\pi j\frac{d_r}{\lambda} (\sqrt{M} - 1) \cos \theta_2 \cos \psi \, e^{2\pi j\frac{d_r}{\lambda} (\sqrt{M} - 1) \sin \theta_2 \cos \psi}
    \end{bmatrix}.
    \end{align}
\end{table*}
\subsection{Sensing Model for Bistatic Radar}

For the sensing object of the bistatic radar setup, the reflected signal is given by:
\begin{equation}
\label{eqn:eq32}
\boldsymbol{y_{sb}}=\gamma_{4}\mathbf{A_{4}}\Theta\h?{b-r}\xs  + \gamma_{3}\mathbf{A_{3}}\xs +\boldsymbol{\zs_{sb}}.
\end{equation}
Here, $\h?{b-r} \in \C^{M \times N_{t}}$  is the channel matrix from the BS to the RIS as previously. The matrix $\Theta$ is the phase shift matrix of RIS. The scalar $\boldsymbol{\zs_{sb}} \in \C^{N_{r}}$ denotes the additive white Gaussian noise (AWGN) with zero mean and variance 
$\sigma_{sb}^2$ for each. The scalar $\gamma_{3} \in \C$ represents the reflection coefficient, accounting for the round-trip path loss and radar cross-section for the BS-to-sensing object path, while $\gamma_{4} \in \C$ denotes the reflection coefficient corresponding to the RIS-to-sensing object link. 
The matrix $\mathbf{A_{3}} \in \C^{N_{r} \times N_{t}}$ is given by:
\begin{equation}
\label{eqn:eq33}
    \mathbf{A_{3}}=\bs_{b}(\phi_{2})\as_{b}(\phi_{1})^\tr,
\end{equation}
where $\as_{b}(\phi_{1}) \in \C^{N_{t}}$, and $\bs_{b}(\phi_{2}) \in \C^{N_{r}}$ represent the transmit and receive array response vectors respectively corresponding to the BS. Unlike the previous case, these vectors are associated with different angles. The angle $\phi_{1}$ denotes the angle of the sensing object related to the transmit antenna, and $\phi_{2}$ refers to the angle of the sensing object related to the receive antenna. The  matrix $\mathbf{A_{4}} \in \C^{N_{r} \times M}$ is given by:
\begin{equation}
\label{eqn:eq34}
\mathbf{A_{4}}=\bs_{b}(\phi_{4})\bs_{b}(\phi_{3},\psi)^\tr.
\end{equation}
Here, $\bs_{b}(\phi_{3},\psi) \in \C^{M}$ represents the array response vector corresponding to the RIS to object. The vector $\bs_{b}(\phi_{4})$ denotes the array response vector from the object to receive antenna. 
The angles $\phi_{3}$ and $\phi_{4}$ denote the angle of the sensing object corresponding to RIS and receive antenna. 

Given the presence of four different angles, the estimation vector now differs from that of the monostatic radar case. Using the estimation vector $\vs_{b}=[\phi _{1} \ \phi _{2} \ \gamma _{1} \ \gamma^*_{1} \ \phi _{3} \ \phi _{4} \ \gamma _{2} \ \gamma^*_{2}]^\tr$ and the bistatic radar model defined by equation (\ref{eqn:eq32}), along with equations (\ref{eqn:eq35}) and (\ref{eqn:eq36}), we derive the FI matrix as follows:
\begin{equation}
\label{eqn:eq35}
    \mathbf{F}_{b} = \mathbb{E}\Biggl[\left(\frac{\partial\ln\rho(y_{sb};\vs_{b})}{\partial\vs_{b}}\right)\left(\frac{\partial\ln\rho(y_{sb};\vs_{b})}{\partial\vs_{b}}\right)^\hr\Biggr],
\end{equation}
where
\begin{align}
\label{eqn:eq36}
   \ln\rho(y_{sb};\vs) &= -N_{t}\ln(\pi\sigma^2_{sb}) \notag \\ & \quad -\frac{1}{\sigma^2_{s}}(y_{sb}-\gamma_{4}\mathbf{A_{4}}\Theta\h?{b-r}\xs- \gamma_{3}\mathbf{A_{3}}\xs)^\hr\notag \\
    & \quad (y_{sb}-\gamma_{4}\mathbf{A_{4}}\Theta\h?{b-r}\xs- \gamma_{3}\mathbf{A_{3}}\xs).
\end{align}
Here, $\rho(y_{sb},\vs_{b})$ is the conditional pdf of $y_{sb}$ w.r.t. the vector $\vs_{b}$. Consequently, FI matrix $\mathbf{F}_{b}$ is given by
\begin{equation}
\label{eqn:eq37}
    \mathbf{F}_{b} = 
    \begin{bmatrix}
    F_{11}  & F_{12}  & F_{13}  & F_{14}  & \quad F_{15}  & F_{16} & F_{17}  & F_{18}\\
    F_{21} &  F_{22} & F_{23} & F_{24} & F_{25} & F_{26} & F_{27}  & F_{28} \\
    F_{31} & F_{32} & F_{33} & 0 & F_{35} & F_{36} & F_{37} & 0 \\
    F_{41} & F_{42} & 0 & F_{44} & F_{45} & F_{46} & 0 & F_{48} \\
    F_{51}  & F_{52}  & F_{53}  & F_{54}  & \quad F_{55}  & F_{56} & F_{57}  & F_{58}\\
    F_{61} &  F_{62} & F_{63} & F_{64} & F_{65} & F_{66} & F_{67}  & F_{68} \\
    F_{71} & F_{72} & F_{73} & 0 & F_{75} & F_{76} & F_{77} & 0 \\
    F_{81} & F_{82} & 0 & F_{84} & F_{85} & F_{86} & F_{87} & F_{88}
    
    \end{bmatrix}.
     \vspace{3pt}
    \end{equation}
The diagonal elements are given by:
   \begin{equation}
   \label{eqn:eq38}
            F_{11} = \frac{2\gamma_{1}\gamma_{1}^*}{\sigma^2_{sb}}\ \text{Tr}\left(\frac{\partial\mathbf{A_{3}}}{\partial\phi_{1}}\mathbf{R_{\xs}}\frac{\partial\mathbf{A_{3}}}{\partial\phi_{1}}^\hr\right)
\end{equation}
 \begin{equation}
   \label{eqn:eq39}
            F_{22} = \frac{2\gamma_{1}\gamma_{1}^*}{\sigma^2_{sb}}\ \text{Tr}\left(\frac{\partial\mathbf{A_{3}}}{\partial\phi_{2}}\mathbf{R_{\xs}}\frac{\partial\mathbf{A_{3}}}{\partial\phi_{2}}^\hr\right)
            \end{equation}
\begin{equation}
\label{eqn:eq40}
        F_{33} =\frac{1}{\sigma^2_{sb}} \text{Tr}\left(\mathbf{A_3}\mathbf{R_{\xs}}\mathbf{A^\hr_3}\right)
        \end{equation}
        \begin{equation}
        \label{eqn:eq41}
            F_{44} = \frac{1}{\sigma^2_{sb}}\text{Tr}\left(\mathbf{A_3}\mathbf{R_{\xs}}\mathbf{A^\hr_3}\right)
        \end{equation}
        \begin{equation}
        \label{eqn:eq42}
            F_{55} = \frac{2\gamma_{2}\gamma_{2}^*}{\sigma^2_{sb}}\ \text{Tr}\left(\frac{\partial\mathbf{A_{4}}}{\partial\phi_{3}}\Theta\h?{b-r}\mathbf{R_{\xs}}\h?{b-r}^\hr\Theta^\hr\frac{\partial\mathbf{A_{4}}}{\partial\phi_{3}}^\hr\right)
        \end{equation}
        \vspace{5pt}
        \begin{equation}
        \label{eqn:eq43}
            F_{66} = \frac{2\gamma_{2}\gamma_{2}^*}{\sigma^2_{sb}}\ \text{Tr}\left(\frac{\partial\mathbf{A_{4}}}{\partial\phi_{4}}\Theta\h?{b-r}\mathbf{R_{\xs}}\h?{b-r}^\hr\Theta^\hr\frac{\partial\mathbf{A_{4}}}{\partial\phi_{4}}^\hr\right)
        \end{equation}
        \begin{equation}
         \label{eqn:eq44}
            F_{77} =\frac{1}{\sigma^2_{sb}}\text{Tr}\left(\mathbf{{A_4}}\Theta\h?{b-r}\mathbf{R_{\xs}}\h?{b-r}^\hr\Theta^\hr\mathbf{{A^\hr_4}}\right)
           \end{equation}
           \vspace{5pt}
        \begin{equation}
        \label{eqn:eq45}
            F_{88} = \frac{1}{\sigma^2_{sb}}\text{Tr}\left(\mathbf{{A_4}}\Theta\h?{b-r}\mathbf{R_{\xs}}\h?{b-r}^\hr\Theta^\hr\mathbf{{A^\hr_4}}\right).
        \end{equation}
\vspace{3pt}
The partial derivatives of $\mathbf{A_{3}}$ are given by 
$\frac{\partial\mathbf{A_{3}}}{\partial\phi_{1}}=\bs_{b}(\phi_2)\dot{\as_{b}}(\phi_1)^\tr$ and $\frac{\partial\mathbf{A_{3}}}{\partial\phi_{2}}=\dot{\bs_{b}}(\phi_2){\as_{b}}(\phi_1)^\tr$. In addition, the partial derivative of $\mathbf{A_{4}}$ w.r.t. $\phi_3$ is $\frac{\partial\mathbf{A_{4}}}{\partial\phi_{3}}=\bs_{b}(\phi_{4})\dot{\bs_{b}}(\phi_{3},\psi)^\tr$ and w.r.t. $\phi_4$ is $\frac{\partial\mathbf{A_{4}}}{\partial\phi_{4}}=\dot{\bs_{b}}(\phi_{4})\bs_{b}(\phi_{3},\psi)^\tr$.

The array response vectors $\as_{b}(\phi_1)$, $\bs_{b}(\phi_2)$  and $\bs_{b}(\phi_4)$ correspond to ULAs and are expressed using the same equation as in (\ref{eqn:e26}), but with the respective new angles. Similarly, the partial derivatives of these array response vectors follow the same form as in (\ref{eqn:e27}), with the replaced angles $\phi_1$, $\phi_2$, and $\phi_4$.
Moreover, the vector $\bs_{b}(\phi_{3},\psi)$ corresponds to a UPA and is represented by the same equation as in (\ref{eqn:eq28}) with the updated angle. The corresponding partial derivative follows the form of equation (\ref{eqn:eq29}).
\section{Effect of Self-interference Through Quantization}

In monostatic radar configurations, where the transmit and receive antennas are co-located, self-interference is significant and cannot be neglected. Therefore, it is essential to account for its impact in system analysis. In this section, we present a method to examine self-interference effect through the quantization of the reflected sensing signal. Unlike existing approaches that directly subtract self-interference, our method retains it and proceeds with quantization to explicitly demonstrate its effect. Consider the reflected sensing signal, including the self-interference, as follows:
\begin{equation}
\label{eqn:eq5_SI}
\boldsymbol{y_{m}}=\gamma_{2}\h?{r-b}\Theta^\tr\mathbf{A_{2}}\Theta\h?{b-r}\xs  + \gamma_{1}\mathbf{A_{1}}\xs+\h?{SI}\xs +\boldsymbol{\zs_{sm}}.
\end{equation}
Here, the matrix $\h?{SI}\in\C^{N_{r} \times N_{t}}$ represents the self-interference between the transmit and receive antennas. As mentioned, we do not subtract the self-interference from $\boldsymbol{y_{m}}$, instead, the signal is quantized directly as follows:
\begin{equation}
\label{eqn:eqQuan}
\boldsymbol{y_{q}}=Q\left(\boldsymbol{y_{m}}\right),
\end{equation}
where $Q$ denotes the quantization operator.
Quantization inherently causes some information loss, leading to performance degradation even in the absence of self-interference. When self-interference is present, this degradation is more pronounced and further amplified. By retaining self-interference during quantization, our method explicitly highlights its impact on the reflected sensing signal, providing a clear measure of how self-interference affects system performance. This approach offers a novel way to characterize self-interference compared to existing state-of-the-art methods. 

The amount of information loss depends on the characteristics of the quantizer, such as the number of bits used. For a quantizer with 
$b$ bits, there are $2^b$ quantization levels. Thereby, for a uniform symmetric mid-riser quantizer, the quantization levels are given by \cite{5456454}:
\begin{equation}
\label{eqn:Qlower}
L_i^\mathrm{lower} =
\begin{cases} 
L_i - \frac{\Delta}{2}, & \text{if } L_i \ge -\frac{\Delta}{2}(2^b-1) \\[1mm]
-\infty, & \text{otherwise}
\end{cases}
\end{equation}

\begin{equation}
\label{eqn:Qupper}
L_i^\mathrm{upper} =
\begin{cases} 
L_i + \frac{\Delta}{2}, & \text{if } L_i \le \frac{\Delta}{2}(2^b-1) \\[1mm]
+\infty, & \text{otherwise}.
\end{cases}
\end{equation}
Once these bounds are defined, the Fisher Information Matrix (FIM) for the quantized sensing signal can be expressed as \cite{5456454}
\begin{equation}
\label{eqn:QFI}
\mathbf{F}_{Q}= \sum_{N_{r},L_i} 
\frac{
\Big(e^{-\frac{(L_i^\mathrm{upper}-\boldsymbol{y_{m}})^2}{2\sigma_m^2}} - e^{-\frac{(L_i^\mathrm{lower}-\boldsymbol{y_{m}})^2}{2\sigma_m^2}}\Big)^2
\frac{\partial \boldsymbol{y_{m}}}{\partial \vs_{m}} \frac{\partial \boldsymbol{y_{m}}}{\partial \vs_{m}}^\hr
}
{2 \pi \sigma_m^2 \Big[ \Phi\Big(\frac{L_i^\mathrm{upper}-\boldsymbol{y_{m}}}{\sigma_m}\Big) - \Phi\Big(\frac{L_i^\mathrm{lower}-\boldsymbol{y_{m}}}{\sigma_m}\Big) \Big]}.
\end{equation}
Here, $N_r$ is the number of receive antennas, $\vs_{m}$ is the parameter vector to be estimated in the monostatic radar, $\sigma_m^2$ is the noise variance per antenna component, and $\Phi(x)$ is the cumulative distribution function of a standard Gaussian. This equation can be used to analyze the sensing performance of the quantized signal. 

Based on the above approach, in  Section VI, through numerical results, we will illustrate the effect of self-interference. 
\section{Problem Formulation and Proposed Solution}

\subsection{Problem Formulation}
This subsection provides the optimization problem formulation focusing on maximizing the weighted sum of the MI and FI \cite{10615441}, which represents a multi-objective optimization problem, subject to a power budget constraint and unimodular constraints on reflection coefficients \cite{10096563,10417011}. The optimization problem can be given as follows:
\begin{subequations}
\label{eqn:op}
\begin{align}
\ds\argmax_{\mathbf{R_{\xs}},\mathbf{\Upsilon}} \quad &  \alpha\text{Tr}(\mathbf{F})+(1-\alpha)\mathrm{MI}\\
    \text{subject to}  \quad
    \label{eqn:sp_1}
    & C1 :\text{Tr}(\mathbf{R_{\xs}}) \leq P, \\
    \label{eqn:sp_2}
    & C2 :\mathbf{R_{\xs}} \ge 0, \\
    \label{eqn:sp_3}
    & C3 : \upsilon_{im}=0, \quad i \neq m,\\
    \label{eqn:sp_4}
   & C4 : |\upsilon_{ii}|=1, \quad i=1,2, \cdots, M.
\end{align}
\end{subequations}

The unimodular constraints on (\ref{eqn:sp_3}) and (\ref{eqn:sp_4}) make the optimization problem  non-convex. Here $\alpha$ is the weight value. The overall optimization problems of the monostatic radar and bistatic radar configurations are given by equations (\ref{eqn:subproblem_2_MS}) and (\ref{eqn:subproblem_2_BS}).
\begin{table*}
\normalsize
\hrule 
\vspace{10pt}
\begin{subequations}
\label{eqn:subproblem_2_MS}
\begin{align}
   \ds\argmax_{\mathbf{R_{\xs}}, \mathbf{\Upsilon}} \quad & \frac{2\alpha\gamma_{1}\gamma_{1}^*}{\sigma_{s}^2}\ \text{Tr}\left(\mathbf{\dot{A_1}}\mathbf{R_{\xs}}\mathbf{\dot{A^\hr_1}}\right) + \frac{2\alpha}{\sigma_{s}^2}\text{Tr}\left(\mathbf{A_1}\mathbf{R_{\xs}}\mathbf{A^\hr_1}\right) + \frac{2\alpha\gamma_{2}\gamma_{2}^*}{\sigma_{s}^2}\ \text{Tr}\left(\h?{r-b}\Theta^\tr\mathbf{\dot{A_2}}\Theta\h?{b-r}\mathbf{R_{\xs}}\h?{b-r}^\hr\Theta^\hr\mathbf{\dot{A^\hr_2}}(\h?{r-b}\Theta^\tr)^\hr\right) \notag \\
& + \frac{2\alpha}{\sigma_{s}^2}\text{Tr}\left(\h?{r-b}\Theta^\tr\mathbf{{A_2}}\Theta\h?{b-r}\mathbf{R_{\xs}}\h?{b-r}^\hr\Theta^\hr\mathbf{{A^\hr_2}}(\h?{r-b}\Theta^\tr)^\hr\right) + (1 - \alpha)\log_{2}\left(1+\frac{\hs^\hr\mathbf{R_{\xs}}\hs}{\sigma_{c}^2} \right) \\
\nonumber \\
    \text{subject to}  \quad
    \label{eqn:sp_1_MS}
    & \text{Tr}(\mathbf{R_{\xs}}) \leq P, \\
    \label{eqn:sp_2_MS}
    & \mathbf{R_{\xs}} \ge 0, \\
     \label{eqn:sp_3_MS}
    & \upsilon_{im}=0, \quad i \neq m,\\
     \label{eqn:sp_4_MS}
    & |\upsilon_{ii}|=1, \quad i=1,2,\cdots, M.
\end{align}
\end{subequations}
\hrule
\end{table*}
\begin{table*}
\normalsize
\vspace{5pt}
\begin{subequations}
\label{eqn:subproblem_2_BS}
\begin{align}
   \ds\argmax_{\mathbf{R_{\xs}},\mathbf{\Upsilon}} \quad & \frac{2\alpha\gamma_{1}\gamma_{1}^*}{\sigma^2_{sb}}\ \text{Tr}\left(\frac{\partial\mathbf{A_{3}}}{\partial\phi_{1}}\mathbf{R_{\xs}}\frac{\partial\mathbf{A^\hr_{3}}}{\partial\phi_{1}}\right)+\frac{2\gamma_{1}\gamma_{1}^*}{\sigma^2_{sb}}\ \text{Tr}\left(\frac{\partial\mathbf{A_{3}}}{\partial\phi_{2}}\mathbf{R_{\xs}}\frac{\partial\mathbf{A^\hr_{3}}}{\partial\phi_{2}}\right)+\frac{2}{\sigma^2_{sb}} \text{Tr}\left(\mathbf{A_3}\mathbf{R_{\xs}}\mathbf{A^\hr_3}\right)\nonumber \\ & + \frac{2\gamma_{2}\gamma_{2}^*}{\sigma^2_{sb}}\ \text{Tr}\left(\frac{\partial\mathbf{A_{4}}}{\partial\phi_{3}}\Theta\h?{b-r}\mathbf{R_{\xs}}\h?{b-r}^\hr\Theta^\hr\frac{\partial\mathbf{A^\hr_{4}}}{\partial\phi_{3}}\right)+\frac{2\gamma_{2}\gamma_{2}^*}{\sigma^2_{sb}}\ \text{Tr}\left(\frac{\partial\mathbf{A_{4}}}{\partial\phi_{4}}\Theta\h?{b-r}\mathbf{R_{\xs}}\h?{b-r}^\hr\Theta^\hr\frac{\partial\mathbf{A^\hr_{4}}}{\partial\phi_{4}}\right)\nonumber \\ & + \frac{2}{\sigma^2_{sb}}\text{Tr}\left(\mathbf{{A_4}}\Theta\h?{b-r}\mathbf{R_{\xs}}\h?{b-r}^\hr\Theta^\hr\mathbf{{A^\hr_4}}\right)+(1 - \alpha)\log_{2}\left(1+\frac{\hs^\hr\mathbf{R_{\xs}}\hs}{\sigma_{c}^2} \right) \\
   \nonumber \\
    \text{subject to}  \quad
    \label{eqn:sp_1_BS}
    & \text{Tr}(\mathbf{R_{\xs}}) \leq P, \\
    \label{eqn:sp_2_BS}
    & \mathbf{R_{\xs}} \ge 0, \\
    \label{eqn:sp_3_BS}
    & \upsilon_{im}=0, \quad i \neq m,\\
    \label{eqn:sp_4_BS}
    & |\upsilon_{ii}|=1, \quad i=1,2,\cdots, M.
\end{align}
\end{subequations}
\hrule
\end{table*}
For simplicity, the optimization problem is decomposed into two subproblems using the Alternating Optimization (AO) method. After dividing the problem into two subproblem the resultant optimization problem is given by \\
For a given $\mathbf{\Upsilon}$:
\begin{subequations}
\label{eqn:subproblem_1}
\begin{align}
   \text{SP1:} \quad \ds\argmax_{\mathbf{R_{\xs}}} \quad & \alpha\text{Tr}(\mathbf{F}) + (1 - \alpha)\mathrm{MI} \\
    \text{subject to}  \quad
    \label{eqn:subproblem_11}
    & \text{Tr}(\mathbf{R_{\xs}}) \leq P, \\
    \label{eqn:subproblem_12}
    & \mathbf{R_{\xs}} \ge 0.
\end{align}
\end{subequations}
For a given $\mathbf{R_{\xs}}$:
\begin{subequations}
\label{eqn:subproblem_2}
\begin{align}
   \text{SP2:} \quad \ds\argmax_{\mathbf{\Upsilon}} \quad & \alpha\text{Tr}(\mathbf{F}) + (1 - \alpha)\mathrm{MI} \\
    \text{subject to}  \quad
    & \upsilon_{im}=0, \quad i \neq m,\\
    & |\upsilon_{ii}|=1, \quad i=1,2,\cdots, M.
\end{align}
\end{subequations}
The first subproblem is a convex optimization problem while the second subproblem is a nonconvex optimization problem. 

\subsection{Proposed Solution}

This subsection presents a solution method for both monostatic and bistatic radar configurations, taking into account a physically consistent model. Furthermore, a solution for the conventional model is also provided to facilitate a comprehensive comparison of the results. For both setups, the first subproblem (SP1) is solved using MATLAB CVX optimization tool, designed to address convex optimization problems. The second subproblem (SP2), on the other hand, is solved using the projected gradient ascent method.

Consider the SP2 of optimizing the reflection coefficient.
Let the objective function, $f(\mathbf{\Upsilon})=\alpha\text{Tr}(\mathbf{F}) + (1 - \alpha)\mathrm{MI}$. We then take the derivative of the objective function with respect to the $\mathbf{\Upsilon}$ as follows:
\begin{equation}
\label{eq:toderi}
\mathbf{G}=\diff{{f}(\mathbf{\Upsilon})}{\mathbf{\Upsilon}}=\alpha\diff{\text{Tr}(\mathbf{F})}{\mathbf{\Upsilon}}+(1 - \alpha)\diff{\mathrm{MI}}{\mathbf{\Upsilon}}.
\end{equation}
In this case, we vary the weight parameter $\alpha$ and take the values in between 0 and 1 with 0.1 intervals. Gradients are calculated for each variation of $\alpha$, and the average gradient is then computed. Subsequently, the diagonal values are extracted from the computed average gradient ensuring that the constraint (\ref{eqn:sp_3_MS}) is satisfied.

\begin{equation}
\label{eqn:eqdiag}
\nabla \mathbf{g}=\tn{Diag}\bigl(\mathbf{G}).
\end{equation}
Following the diagonal extraction step, the phase shift vector is updated. The equation below gives the key step of the gradient ascent  method:
\begin{equation}
    \bm\upsilon=\bm\upsilon+\mu\nabla \mathbf{g}.
\end{equation}
Here, $\mu$ is a fix step size. Then we take the projection to satisfy the constraint as follows:
\begin{equation}
\bm\upsilon =\frac{\bm\upsilon}{\left|{\bm\upsilon}\right|}.
\end{equation}
In the context of monostatic radar, the derivative of $\text{Tr}(\mathbf{F})$ corresponding to the physically consistent model is expressed as in the equation (\ref{eq:phy-F}) and the equation (\ref{eq:con-F}) gives the derivative of $\text{Tr}(\mathbf{F})$ corresponding to the conventional model.  On the other hand, in the context of bistatic radar, the derivatives of $\text{Tr}(\mathbf{F})$ for the physically consistent and conventional models are given by equations (\ref{eqn:eqn_bi_deriv_phy}) and (\ref{eqn:eqnbi_deri_con}), respectively.
Moreover, for both setups, the derivatives of MI of the physically consistent and conventional models are given by the corresponding equations (\ref{eq:phy-MI}) and (\ref{eq:con-MI}), respectively. The overall algorithm for the monostatic radar considering physically consistent model is given in \textbf{Algorithm \ref{algo:joint_op}}. 
\begin{algorithm}
\caption{Proposed Algorithm for monostatic radar considering physically-consistent model}
\label{algo:joint_op}

\begin{algorithmic}[1]
\renewcommand{\algorithmicrequire}{\textbf{Input:}}
\renewcommand{\algorithmicensure}{\textbf{Output:}}
\REQUIRE The channel matrices $\h?{b-r}$, $\h?{r-u}$, $\mathbf{A_{1}}$, $\mathbf{A_{2}}$, $\mathbf{S}$.
\vspace{1pt}
\STATE Initialize $\bm\upsilon_{0}$.
\vspace{1pt}
\STATE Set weight value $\alpha=0:0.1:1$
\STATE Set $\mathbf{\Upsilon}_{1}=\tn{Diag}(\bm\upsilon_{0})$
\REPEAT 
\FOR{$k=1,2,\ldots,length(\alpha)$}
\FOR{$CVX_{iter}=1,2,\ldots,C$}
\STATE Compute $\mathbf{R_{\xs}}_{k}$ using CVX algorithm.  
\ENDFOR
\STATE Compute $\diff{\text{Tr}(\mathbf{F})}{\mathbf{\Upsilon}_{k}}$ using \eqref{eq:phy-F} and $\diff{\mathrm{MI}}{\mathbf{\Upsilon}_{k}}$ \eqref{eq:phy-MI}.  
\STATE Compute $\mathbf{G}_{k}$ using \eqref{eq:toderi}.
\vspace{1.5pt}
\ENDFOR
\STATE Compute $\mathbf{G}_{all}=\Sigma_{k}\mathbf{G}_{k}$.
\vspace{1.5pt}
\STATE Compute $\mathbf{G}_{avg}=\frac{\mathbf{G}_{all}}{length(\alpha)}$.
\vspace{1.5pt}
\STATE Compute $\nabla \mathbf{g}=\tn{Diag}\bigl(\mathbf{G}_{avg})$.
\vspace{1.5pt}
\STATE Compute $\bm\upsilon=\bm\upsilon+\mu\nabla \mathbf{g}$.
\vspace{1.5pt}
\STATE Compute $\bm\upsilon=\frac{\bm\upsilon}{\abs{\bm\upsilon}}$.
\UNTIL
\RETURN Optimum $\mathbf{R_{\xs}}$, $\mathbf{\Upsilon}$.
\end{algorithmic}
\end{algorithm}
\begin{table*}
\normalsize
\hrule
\vspace{7pt}
    \begin{align}
\label{eq:phy-F}
     &\diff{\text{Tr}(\mathbf{F})}{\mathbf{\Upsilon}}=  \nonumber\frac{2\gamma_{2}\gamma_{2}^*}{\sigma^2_{s}}\Bigl(\h?{r-b}^\hr\h?{r-b}(\mathbf{\Upsilon}^{-1}-\mathbf{S})^{-1}\mathbf{\dot{A_2}}(\mathbf{\Upsilon}^{-1}-\mathbf{S})^{-1}\h?{b-r} 
  \mathbf{R_{\xs}\h?{b-r}^\hr}(\mathbf{\Upsilon}^{-1}-\mathbf{S})^{-\hr}\mathbf{\Upsilon}^{-\hr}(\mathbf{\Upsilon}^{-1}-\mathbf{S})^{-\hr}\mathbf{\dot{A^\hr_2}}
  (\mathbf{\Upsilon}^{-1}-\mathbf{S})^{-\hr}\mathbf{\Upsilon}^{-\hr}\nonumber \\ & +\mathbf{\dot{A^\hr_2}}(\mathbf{\Upsilon}^{-1}-\mathbf{S})^{-\hr}\mathbf{\Upsilon}^{-\hr}(\mathbf{\Upsilon}^{-1}-\mathbf{S})^{-\hr}
  \h?{r-b}^\hr\h?{r-b}(\mathbf{\Upsilon}^{-1}-\mathbf{S})^{-1}\mathbf{\dot{A_2}}(\mathbf{\Upsilon}^{-1}-\mathbf{S})^{-1}\h?{b-r}\mathbf{R_{\xs}\h?{b-r}^\hr}(\mathbf{\Upsilon}^{-1}-\mathbf{S})^{-\hr}\mathbf{\Upsilon}^{-\hr}\Bigr) \nonumber \\ & +
  \frac{2}{\sigma^2_{s}}\Bigl(\h?{r-b}^\hr\h?{r-b} 
(\mathbf{\Upsilon}^{-1}-\mathbf{S})^{-1}\mathbf{{A_2}}(\mathbf{\Upsilon}^{-1}-\mathbf{S})^{-1}\h?{b-r} 
  \mathbf{R_{\xs}\h?{b-r}^\hr}(\mathbf{\Upsilon}^{-1}-\mathbf{S})^{-\hr}\mathbf{\Upsilon}^{-\hr}(\mathbf{\Upsilon}^{-1}-\mathbf{S})^{-\hr}\mathbf{{A^\hr_2}}
 (\mathbf{\Upsilon}^{-1}-\mathbf{S})^{-\hr}\mathbf{\Upsilon}^{-\hr}\nonumber \\ &+
 \mathbf{{A^\hr_2}}(\mathbf{\Upsilon}^{-1}-\mathbf{S})^{-\hr}\mathbf{\Upsilon}^{-\hr}(\mathbf{\Upsilon}^{-1}-\mathbf{S})^{-\hr}\h?{r-b}^\hr\h?{r-b}(\mathbf{\Upsilon}^{-1}-\mathbf{S})^{-1}\mathbf{{A_2}}(\mathbf{\Upsilon}^{-1}-\mathbf{S})^{-1}\h?{b-r}\mathbf{R_{\xs}\h?{b-r}^\hr}(\mathbf{\Upsilon}^{-1}-\mathbf{S})^{-\hr}\mathbf{\Upsilon}^{-\hr}  
  \Bigr),
\end{align}
\begin{align}
\label{eq:con-F}
    & \diff{\text{Tr}(\mathbf{F})}{\mathbf{\Upsilon}}=  \frac{2\gamma_{2}\gamma_{2}^*}{\sigma^2_{s}}\Bigl(\h?{r-b}^\hr\h?{r-b}\mathbf{\Upsilon}\mathbf{\dot{A_2}}\mathbf{\Upsilon}\h?{b-r}\mathbf{R_{\xs}\h?{b-r}^\hr}\mathbf{\Upsilon}^{\hr}\mathbf{\dot{A^\hr_2}} +\mathbf{\dot{A^\hr_2}}\mathbf{\Upsilon}^{\hr}
  \h?{r-b}^\hr\h?{r-b}\mathbf{\Upsilon}\mathbf{\dot{A_2}}\mathbf{\Upsilon}\h?{b-r}\mathbf{R_{\xs}\h?{b-r}^\hr}\Bigr)
  \nonumber \\ & + 
  \frac{2}{\sigma^2_{s}}\Bigl(\h?{r-b}^\hr\h?{r-b} 
\mathbf{\Upsilon}\mathbf{{A_2}}\mathbf{\Upsilon}\h?{b-r}
  \mathbf{R_{\xs}\h?{b-r}^\hr}\mathbf{\Upsilon}^\hr\mathbf{{A^\hr_2}}+\mathbf{{A^\hr_2}}\mathbf{\Upsilon}^{\hr}\h?{r-b}^\hr\h?{r-b}\mathbf{\Upsilon}\mathbf{{A_2}}\mathbf{\Upsilon}\h?{b-r}\mathbf{R_{\xs}\h?{b-r}^\hr}\Bigr),
\end{align}
\begin{align}
\label{eqn:eqn_bi_deriv_phy}
    & \diff{\text{Tr}(\mathbf{F})}{\mathbf{\Upsilon}}= \frac{4\gamma_{2}\gamma_{2}^*}{\sigma^2_{sm}}\Bigl(\frac{\partial\mathbf{A^\hr_{4}}}{\partial\phi_{3}}\frac{\partial\mathbf{A_{4}}}{\partial\phi_{3}}\mathbf{\Upsilon}^{-\hr}(\mathbf{\Upsilon}^{-1}-\mathbf{S})^{-\hr}((\mathbf{\Upsilon}^{-1}-\mathbf{S})^{-1})^*\h?{b-r}\mathbf{R_{\xs}}\h?{b-r}^\hr(\mathbf{\Upsilon}^{-1}-\mathbf{S})^{-\hr}\mathbf{\Upsilon}^{-\hr}\Bigr)
  \\ & + 
  \frac{4\gamma_{2}\gamma_{2}^*}{\sigma^2_{sm}}\Bigl(\frac{\partial\mathbf{A^\hr_{4}}}{\partial\phi_{4}}\frac{\partial\mathbf{A_{4}}}{\partial\phi_{4}}\mathbf{\Upsilon}^{-\hr}(\mathbf{\Upsilon}^{-1}-\mathbf{S})^{-\hr}((\mathbf{\Upsilon}^{-1}-\mathbf{S})^{-1})^*\h?{b-r}\mathbf{R_{\xs}}\h?{b-r}^\hr(\mathbf{\Upsilon}^{-1}-\mathbf{S})^{-\hr}\mathbf{\Upsilon}^{-\hr}\Bigr) \nonumber \\ & + 
  \frac{4}{\sigma^2_{sm}}\Bigl(\mathbf{A^\hr_{4}}\mathbf{A_{4}}\mathbf{\Upsilon}^{-\hr}(\mathbf{\Upsilon}^{-1}-\mathbf{S})^{-\hr}((\mathbf{\Upsilon}^{-1}-\mathbf{S})^{-1})^*\h?{b-r}\mathbf{R_{\xs}}\h?{b-r}^\hr(\mathbf{\Upsilon}^{-1}-\mathbf{S})^{-\hr}\mathbf{\Upsilon}^{-\hr}\Bigr) \nonumber.
\end{align}
\begin{align}
\label{eqn:eqnbi_deri_con}
    & \diff{\text{Tr}(\mathbf{F})}{\mathbf{\Upsilon}}= \frac{4\gamma_{2}\gamma_{2}^*}{\sigma^2_{sm}}\Bigl(\frac{\partial\mathbf{A^\hr_{4}}}{\partial\phi_{3}}\frac{\partial\mathbf{A_{4}}}{\partial\phi_{3}}{\mathbf{\Upsilon}^*}\h?{b-r}\mathbf{R_{\xs}}\h?{b-r}^\hr\Bigr)+ \frac{4\gamma_{2}\gamma_{2}^*}{\sigma^2_{sm}}\Bigl(\frac{\partial\mathbf{A^\hr_{4}}}{\partial\phi_{4}}\frac{\partial\mathbf{A_{4}}}{\partial\phi_{4}}{\mathbf{\Upsilon}^*}\h?{b-r}\mathbf{R_{\xs}}\h?{b-r}^\hr\Bigr)+
  \frac{4}{\sigma^2_{sm}}\Bigl(\mathbf{A^\hr_{4}}\mathbf{A_{4}}{\mathbf{\Upsilon}^*}\h?{b-r}\mathbf{R_{\xs}}\h?{b-r}^\hr\Bigr),
\end{align}
\begin{align}
\label{eq:phy-MI}
    & \diff{\mathrm{MI}}{\mathbf{\Upsilon}}=
    \frac{2\hs?{r-u}^\hr\mathbf{\Upsilon}^{-\hr}(\mathbf{\Upsilon}^{-1}-\mathbf{S})^{-\hr}\left(\hs?{r-u}^\hr(\mathbf{\Upsilon}^{-1}-\mathbf{S})^{-1}\h?{b-r} + \hs?{b-u}^\hr\right)^*\mathbf{R_{\xs}}\h?{b-r}^\hr(\mathbf{\Upsilon}^{-1}-\mathbf{S})^{-\hr}\mathbf{\Upsilon}^{-\hr}}{\sigma^2_{c}\ln(2)\Bigl(1+\frac{(\hs?{r-u}^\hr(\mathbf{\Upsilon}^{-1}-\mathbf{S})^{-1}\h?{b-r} + \hs?{b-u}^\hr)\mathbf{R_{\xs}}(\hs?{r-u}^\hr(\mathbf{\Upsilon}^{-1}-\mathbf{S})^{-1}\h?{b-r}  + \hs?{b-u}^\hr)^\hr}{\sigma^2_{c}}\Bigr)},
\end{align}
\begin{align}
\label{eq:con-MI}
    \diff{\mathrm{MI}}{\mathbf{\Upsilon}}= 
    \frac{2\hs?{r-u}^\hr\left(\hs?{r-u}^\hr\mathbf{\Upsilon}\h?{b-r} + \hs?{b-u}^\hr\right)^*\mathbf{R_{\xs}}\h?{b-r}^\hr}{\sigma^2_{c}\ln(2)\Bigl(1+\frac{(\hs?{r-u}^\hr\mathbf{\Upsilon}\h?{b-r} + \hs?{b-u}^\hr)\mathbf{R_{\xs}}(\hs?{r-u}^\hr\mathbf{\Upsilon}\h?{b-r}  + \hs?{b-u}^\hr)^\hr}{\sigma^2_{c}}\Bigr)}.
\end{align}
\end{table*}

\subsection{Analysis of Computational Complexity}

Consider \textbf{Algorithm 1} for monostatic radar. Let the outer loop iterate for
$I_{max}$ iterations, while the first inner loop executes for $K=length(\alpha)$.  The second inner loop involves CVX optimization of $\mathbf{R_{\xs}}$ which requires $\mathcal{O}(N_t^2)$ operations, where $N_t$ denotes the number of transmit antennas. As the first inner optimization is repeated $K$ times, the total computational complexity for the CVX process is $K\mathcal{O}(N_t^2)$. Calculating the two gradients requires $\mathcal{O}(M^3)$ operations each, resulting in a total computational complexity of $2K\mathcal{O}(M^3)$. This is the dominant factor, as it involves matrix multiplications and matrix inversion calculations. Then calculating the total gradient only takes $K\mathcal{O}(M)$ operations in total. After the two inner loops, all other steps including the summation, averaging of gradients, diagonal extraction, phase shift vector update, and normalization each require $\mathcal{O}(M)$ operations, resulting in a total complexity of $5\mathcal{O}(M)$. Hence, the overall complexity is $I_{max}\left(K\left(\mathcal{O}(N_t^2)+2\mathcal{O}(M^3)+\mathcal{O}(M)\right)+5\mathcal{O}(M)\right)$.
\section{Numerical Results}

\subsection{Simulation Parameters}

We use parametric channel models for our simulations and assume a total transmit power of $P=30$ dBm. The path-loss at the reference distance is defined as $C_0 = -20$ dB, and the noise variance for the communication model is $\sigma^2_c = 50$ dBm. The sensing parameters are the same as those provided in \cite{10615441}. Far-field approximations are applied in the simulations, with a frequency of 3 GHz and $N_t=N_r=4$. To ensure efficient sensing, the rank of the matrix $\h?{b-r}$ needs to be greater than 1. To achieve this, we place the RIS 25m from the BS while increasing the number of multipath components between the BS and the RIS. Here, we consider 15 multipath components. Also, we assume that the sensing object is at $0^\circ$ angle.  
For the simulations, the weight value $\alpha$ varies from 0 to 1 in increments of 0.1. 
For benchmarking, we use two approaches: (a) a physically-consistent model, and (b) a conventional model that ignores mutual coupling. 

\subsection{Simulation Results for Monostatic Radar Configuration}
\begin{figure}[h!]
\centering
\includegraphics[scale=0.5]{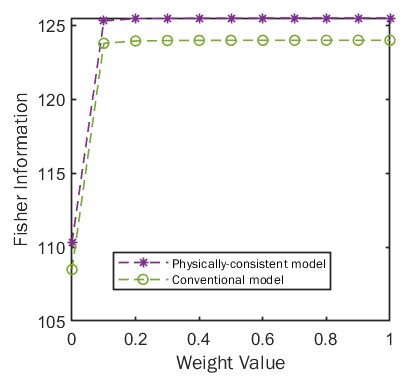}
\caption{Variation of FI with weight value for $M=100$ for monostatic radar.}
\label{fig:FI_wv}
\end{figure}

\begin{figure*}
\centering
\begin{subfigure}{.333\textwidth}
\centering
\includegraphics[scale=0.4]{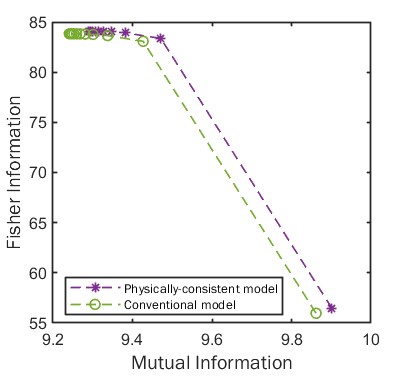}
\caption{\footnotesize $M=64$.}
\label{fig:sumrate_64_1}
\end{subfigure}%
\begin{subfigure}{.334\textwidth}
\centering
\includegraphics[scale=0.4]{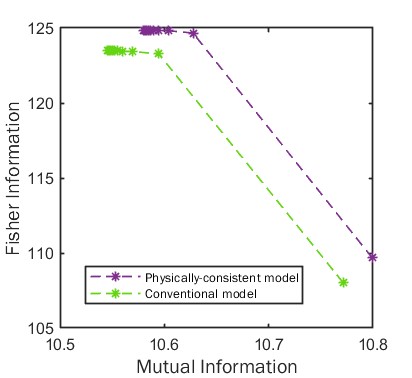}
\caption{\footnotesize $M=100$.}
\label{fig:sumrate_64_2}
\end{subfigure}%
\begin{subfigure}{.333\textwidth}
\centering
\includegraphics[scale=0.4]{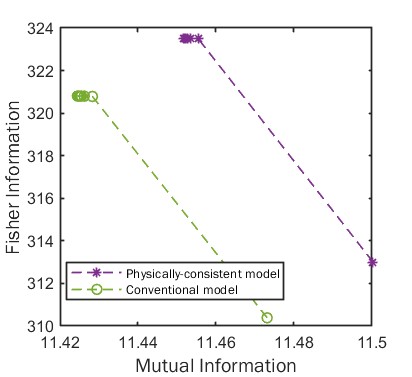}
\caption{\footnotesize $M=144$.}
\label{fig:sumrate_64_3}
\end{subfigure}
\caption{Trade-off between FI and MI for different number of RIS elements with element spacing of $\lambda/2$ for monostatic radar.}
\label{fig:sumrate_RIS}
\end{figure*}

\begin{figure}[h!]
\centering
\includegraphics[scale=0.42]{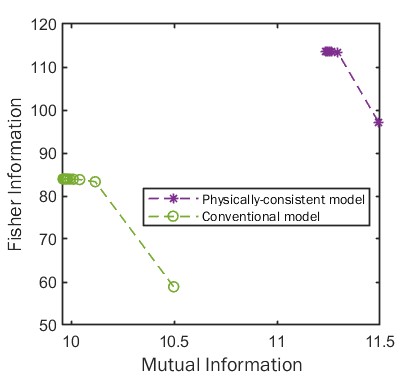}
\caption{Trade-off between FI and MI with  $M=64$ and element spacing of $\lambda/4$ for monostatic radar.}
\label{fig:FI_spacing}
\end{figure}

Fig. \ref{fig:FI_wv} illustrates the variation of FI with the weight value, considering 100 RIS elements and an element spacing of 
$\lambda/2$.  It can be observed that when mutual coupling is considered in the optimization, the FI increases compared to the conventional model as the weight value increases. Fig. \ref{fig:sumrate_RIS} shows the trade-off between FI and MI. As observed, the gap between the conventional and physically-consistent models widens with an increasing number of RIS elements. This indicates that, while expanding the surface area and maintaining the same spacing between consecutive elements, the impact of mutual coupling becomes increasingly significant. Fig. \ref{fig:FI_spacing} illustrates the trade-off between FI and MI with an element spacing of $\lambda/4$ and $M=64$. Compared to Fig. \ref{fig:sumrate_64_1}, which illustrates 64 elements with a spacing of $\lambda/2$, a considerable increase is observed when the spacing is reduced. This observation highlights that mutual coupling becomes crucial for both sensing and communication when the RIS elements become closer. Ignoring mutual coupling, especially  at a spacing of $\lambda/4$ or less, leads to significant performance degradation. Higher performance improvements and more insights can be observed in a multi-user communication setting, where the effects of mutual coupling become even more pronounced. 

\subsection{Simulation Results for Bistatic Radar Configuration}

\begin{figure}[h!]
\centering
\includegraphics[scale=0.42]{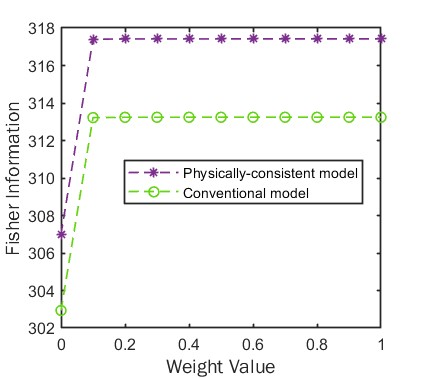}
\caption{Variation of FI with weight value for $M=100$ for bistatic radar.}
\label{fig:FI_wvbi}
\end{figure}

\begin{figure*}
\centering
\begin{subfigure}{.333\textwidth}
\centering
\includegraphics[scale=0.425]{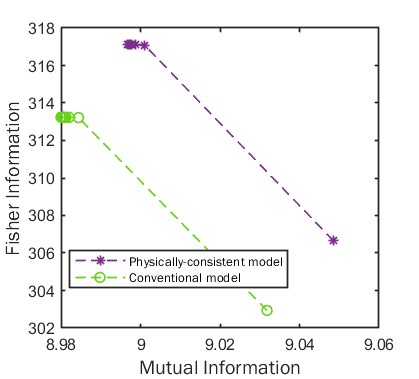}
\caption{\footnotesize $M=64$.}
\label{fig:sumrate_64_1}
\end{subfigure}%
\begin{subfigure}{.334\textwidth}
\centering
\includegraphics[scale=0.425]{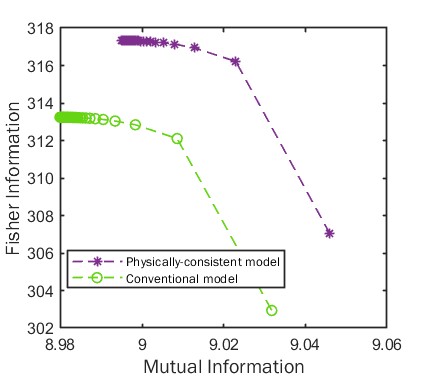}
\caption{\footnotesize $M=64$ with weight interval 0.01.}
\label{fig:bi_weight_large}
\end{subfigure}%
\begin{subfigure}{.333\textwidth}
\centering
\includegraphics[scale=0.425]{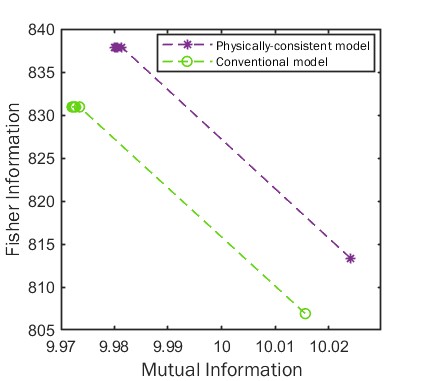}
\caption{\footnotesize $M=100$.}
\label{fig:sumrate_64_3}
\end{subfigure}
\caption{Trade-off between FI and MI for different number of RIS elements with  element spacing of $\lambda/2$ for bistatic radar considering AOD $2\pi/3$.}
\label{fig:bistatic_2pi3}
\end{figure*}

Figure \ref{fig:FI_wvbi} shows the variation of FI with respect to the weight value. In the bistatic radar configuration, the physically-consistent model also outperforms the conventional model. Figure \ref{fig:bistatic_2pi3} presents the trade-off between FI and MI for the bistatic radar configuration when the Angle of Departure (AOD) is $2\pi/3$. Similar to the monostatic case, the physically-consistent model again outperforms the conventional model. However, compared to the monostatic configuration, the performance gap between the two models is larger in the bistatic case, highlighting the reduced self-interference of the bistatic configuration. Figure \ref{fig:bi_weight_large} depicts the trade-off between FI and MI when the weight interval $\alpha$ is varied in steps of 0.01 instead of 0.1. Reducing the interval results in smoother curves, but significantly increases the simulation time. For this reason, for the other results we use a weight interval of 0.1. 

\begin{figure*}
\vspace{-10pt}
\centering
\begin{subfigure}{.5\textwidth}
\centering
\includegraphics[scale=0.465]{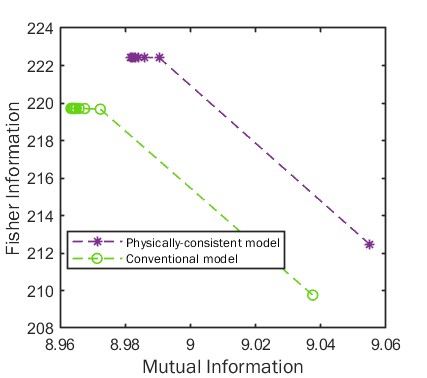}
\caption{\footnotesize $M=64$.}
\label{fig:sumrate_64bi_3pi5}
\end{subfigure}%
\begin{subfigure}{.5\textwidth}
\centering
\includegraphics[scale=0.465]{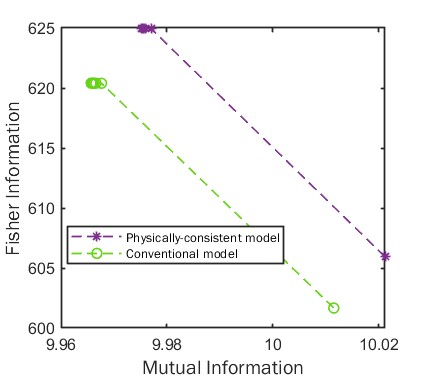}
\caption{\footnotesize $M=100$.}
\label{fig:sumrate_100bi_3pi/5}
\end{subfigure}%
\caption{Trade-off between FI and MI for different number of RIS elements with  element spacing of $\lambda/2$ for bistatic radar considering AOD $3\pi/5$.}
\label{fig:bistatic_3pi5}
\end{figure*}

\begin{figure}[h!]
\centering
\includegraphics[scale=0.45]{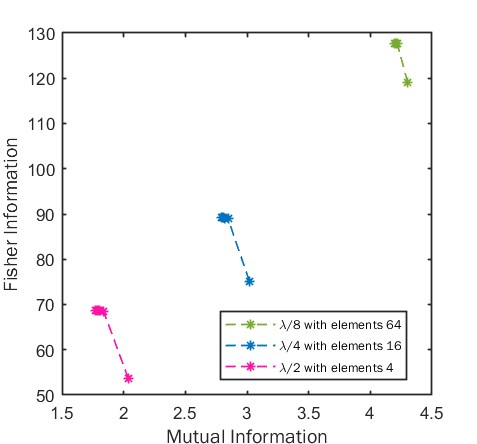}
\caption{Trade-off between FI and MI in the physically-consistent bistatic radar model with AoD $2\pi/3$ for varying element spacing while the surface area remains the same.}
\label{fig:FI_bi_element_spacing}
\end{figure}

Note that, in the bistatic radar configuration, additional degrees of freedom can be achieved. To illustrate this, we present simulation results in Fig.~\ref{fig:bistatic_3pi5} by varying the angle of departure (AoD). Compared to the $2\pi/3$ case, the FI values for both $M=64$ and $M=100$ change when the AoD is set to $3\pi/5$. Thus, varying the AoD directly affects the resulting FI values. This demonstrates that a bistatic radar configuration provides greater flexibility, since adjusting the AoD introduces more variations and thereby increases the available degrees of freedom. Figure \ref{fig:FI_bi_element_spacing} illustrates the trade-off between FI and MI in the physically-consistent model, where the element spacing of the RIS is varied while maintaining a fixed surface area. As shown, placing the elements closer allows more elements to fit within the same surface area. Specifically, for spacings of $\lambda/8$, $\lambda/4$, and $\lambda/2$, the surface accommodates 64, 16, and 4 elements, respectively. The figure shows that closer spacing with more elements yields better performance. This is because reduced spacing enhances the impact of mutual coupling between elements, and having more elements further increases this effect. With proper optimization, these coupling effects can be exploited, leading to improved performance when the elements are placed closer.
\subsection{Impact of Self-interference (SI) Through Quantization}

In this subsection, we analyze the impact of SI on the sensing performance considering physically-consistent model. To model the SI arising between the transmit and receive antenna elements, we adopt the free-space path-loss equation:
\begin{equation}
\label{eqn:Self_in}
    h(p,q) = \left( \frac{\lambda}{4 \pi d(p,q)} \right)^2\, e^{-j k d(p,q)},
\end{equation}
where $p = 1, \ldots, N_r$ and $q = 1, \ldots, N_t$. Here, $h(p,q)$ denotes the $(p,q)$-th element of the self-interference matrix $\mathbf{H}_{\mathrm{SI}}$, and $d(p,q)$ is the distance between the $p$-th receive antenna element and the $q$-th transmit antenna element.  
We first obtain the optimized values of the transmit covariance matrix $\mathbf{R}_{\mathbf{x}}$ and the phase-shifter matrix $\mathbf{\Upsilon}$ for JCAS without considering quantization. Once these optimal values are determined, the sensing signal is quantized, and the FI is calculated using \eqref{eqn:QFI} to evaluate the effect of quantization. The quantization is performed for $b = 1, 3, 4, 8$ bits. The quantization step size for each bit resolution is calculated using the following equation:
\begin{equation}
\label{eqn:quantization_step_size}
    \Delta = \Delta_{\mathrm{opt}} \sqrt{\sigma^2},
\end{equation}
where $\sigma^2$ is the variance of the overall sensing signal. Table~\ref{tab:quantization_steps} lists the values of $\Delta_{\mathrm{opt}}$ and the corresponding maximum quantization step size $\Delta_{\max}$ for each bit resolution \cite{Mezghani2016}.
\begin{table}[h!]
\centering
\caption{Quantization step size for different bit resolutions}
\label{tab:quantization_steps}
\begin{tabular}{c c c}
\hline
Bit resolution $b$ & $\Delta_{\mathrm{opt}}$ & $\Delta_{\max}$ \\ 
\hline
1 & $\sqrt{\frac{8}{\pi}}$ & $\sqrt{\frac{8}{\pi}} \sqrt{\sigma^2}$ \\ 
3 & 0.5860 & $0.5860 \sqrt{\sigma^2}$ \\ 
4 & 0.3352 & $0.3352 \sqrt{\sigma^2}$ \\ 
8 & 0.0308 & $0.0308 \sqrt{\sigma^2}$ \\ 
\hline
\end{tabular}
\end{table}

\begin{figure}[h!]
\centering
\includegraphics[scale=0.42]{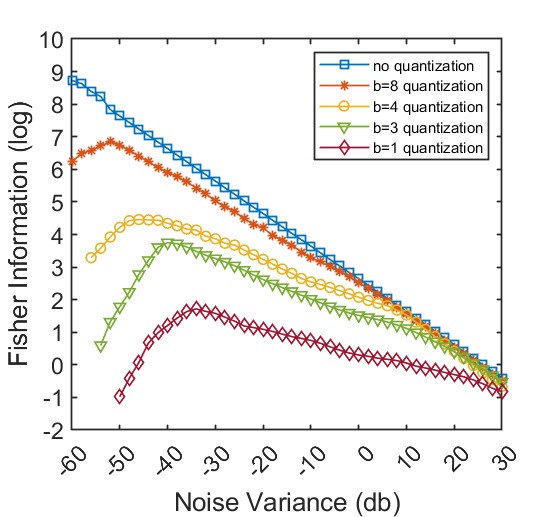}
\caption{FI versus noise variance in the physically-consistent model for different quantization bits without SI present.}
\label{fig:allquantization}
\end{figure}

\begin{figure*}
\vspace{-25pt}
\centering
\begin{subfigure}{.5\textwidth}
\centering
\includegraphics[scale=0.42]{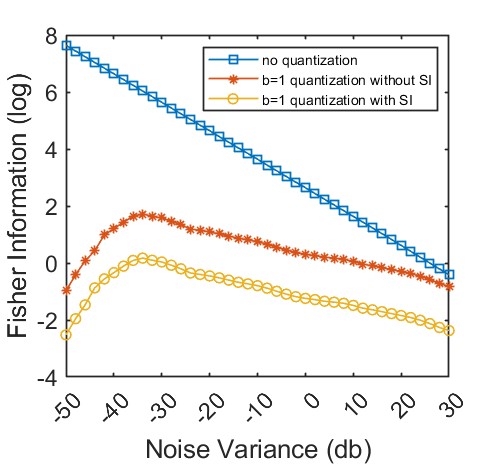}
\caption{\footnotesize $b=1$.}
\label{fig:1bit_qun}
\end{subfigure}%
\begin{subfigure}{.5\textwidth}
\centering
\includegraphics[scale=0.42]{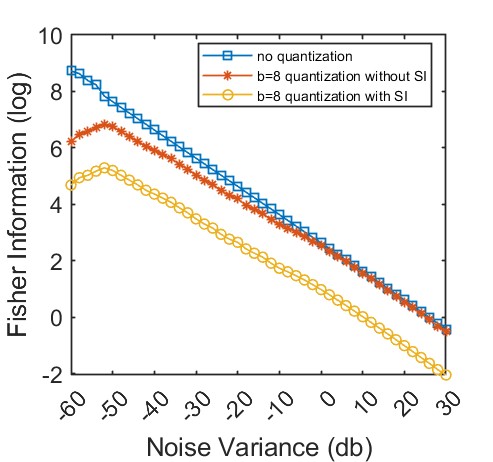}
\caption{\footnotesize $b=8$.}
\label{fig:8bit_quan}
\end{subfigure}%
\caption{FI versus noise variance in the physically-consistent model without SI and with SI present.}
\label{fig:Quan_SI_new}
\end{figure*}

Figure \ref{fig:allquantization} illustrates the variation of FI with noise variance, without considering SI. As observed, higher quantization bit levels result in smaller information loss. The effect of quantization is most pronounced at high SNR (low noise variance), where the FI curves show noticeable deviations from the unquantized case. At low SNR (high noise variance), the FI values are close to those of the unquantized signal. However, when the number of quantization bits is very low (e.g., b = 1, 3), a significant loss of information is observed even at low SNR. This demonstrates that increasing the quantization resolution reduces the degradation in sensing performance caused by quantization. Figure \ref{fig:Quan_SI_new} illustrates the effect of quantization on information loss when SI is present in the system. As shown in Figures \ref{fig:1bit_qun} and \ref{fig:8bit_quan}, corresponding to $b=1$ and $b=8$ quantization bits respectively, the presence of SI leads to a significant loss of information across both low and high SNR levels. This highlights that SI, particularly in a monostatic radar configuration, can substantially degrade the sensing performance.

\section{Conclusion}
We have proposed a novel RIS-assisted joint communication and sensing framework that incorporates the mutual coupling effects in the RIS. For both communication and sensing, with an objective of maximizing the weighted sum of Fisher information and mutual information, we have optimized the  active beamforming at the BS along with the passive beamforming at the  RIS in a physically-consistent environment. The numerical results, based on the Pareto boundary of FI and MI, show that mutual coupling significantly improves performance in both sensing and communication compared to conventional RIS models. One possible extension of the work could be optimization of joint communication and sensing considering stacked RIS, which would provide more degrees of freedom (compared to a single layer RIS) in order to optimize the system performance.

\bibliographystyle{IEEEtran}
\bibliography{bib} 
\end{document}